\documentclass[journal=jacsat,manuscript=article]{achemso}

\usepackage[version=3]{mhchem} 

\usepackage{amsmath}
\usepackage{gensymb,siunitx} 
\usepackage{textcomp}
\usepackage{url} 
\usepackage{threeparttable}
\usepackage{booktabs}
\usepackage{enumitem} 
\usepackage{varioref} 
\usepackage{hyperref}

\usepackage{titlesec}
\titlelabel{\thetitle.\quad}

\author{Yasushi Saka}
\email{y.saka@abdn.ac.uk}
\phone{+44 (0)1224437407}
\author{Murray MacPherson}
\author{Claudiu V. Giuraniuc}
\affiliation[University of Aberdeen]
{School of Medicine, Medical Sciences \& Nutrition, Institute of Medical Sciences, University of Aberdeen, Foresterhill, Aberdeen
AB25 2ZD, United Kingdom}

\title[Dynamic biochemical gradients]
  {Generation and precise control of dynamic biochemical gradients for cellular assays}

\begin{document}
\begin{abstract}

\noindent Spatial gradients of diffusible signalling molecules play crucial roles in controlling diverse cellular behaviour such as cell differentiation, tissue patterning and chemotaxis. 
In this paper, we report the design and testing of a microfluidic device for diffusion-based gradient generation for cellular assays.
A unique channel design of the device eliminates cross-flow between the source and sink channels, thereby stabilising gradients by passive diffusion.
The platform also enables quick and flexible control of chemical concentration that makes highly dynamic gradients in diffusion chambers. 
A model with the first approximation of diffusion and surface adsorption of molecules recapitulates the experimentally observed gradients.
Budding yeast cells cultured in a gradient of a chemical inducer expressed a reporter fluorescence protein in a concentration-dependent manner.
This microfluidic platform serves as a versatile prototype applicable to a broad range of biomedical investigations.

%
\end{abstract}

\newpage
\section{Introduction}

Biochemical gradients are ubiquitous in biological systems. 
Gradients convey spatial and temporal information and function as cues for cellular decisions such as differentiation, proliferation and chemotaxis.
Knowing how cells respond to them is thus fundamentally important to biomedicine and biotechnology applications. 
A well-studied example of biochemical gradients is that of morphogens that pattern embryonic tissues in a concentration-dependent manner (reviewed in \cite{Smith:2009ve,Nahmad:2011zr,Rogers:2011jk,Briscoe:2015ty}).
Recent experimental evidence has suggested that the response of cells to morphogen gradients is regulated dynamically as the gradients change over time \cite{Nahmad:2011zr,Briscoe:2015ty}.
Dynamics of morphogen gradients are therefore crucial for tissue patterning, as corroborated by our theoretical study \cite{batmanov2012self}.
As biochemical gradients \emph{in vivo} may never be static, investigations of cellular response to gradients require experimental means to control gradients in a dynamic manner to reproduce a natural cellular environment.

Microfluidic devices are becoming popular for studying gradients as they can generate and maintain desired gradient profiles stably for a long period \cite{Kim:2010zr,Berthier:2014yq,Lin:2015qd}.
A number of different designs of microfluidic gradient generators have been proposed, which can be grouped into two categories \cite{Lin:2015qd}: flow-based \cite{Jeon:2000aa,Irimia:2006kq,Cooksey:2009uq,Millet:2010kq,Chung:2014fj} and diffusion-based gradient generators \cite{Abhyankar:2006xy,Keenan:2006aa,Irimia:2007fk,Cheng:2007kx,Paliwal:2007aa,Mosadegh:2007fr,Kim:2009ys,Skoge:2010vn,Lee:2012yg,Atencia:2012fk,Frank:2013kq,Beck:2016rt,Uzel:2016aa}.
The flow-based gradient is established by the diffusion at the interface of laminar flows.
Diffusion-based designs utilise a passive diffusion between a source and a sink flow channel or reservoir of chemicals, which generates a linear gradient at steady states.
Their advantages and disadvantages have previously been discussed \cite{Berthier:2014yq,Lin:2015qd}.
For example, flow-based designs, unlike diffusion-based ones, can generate gradients of complex profile and maintain them stably for a prolonged time.
The method of creating such non-linear gradients can reveal unexpected cell behaviour, for instance, in neutrophil chemotaxis \cite{Li-Jeon:2002qf,Irimia:2006kq}.
\newline\indent 
One of the major drawbacks of the flow-based gradient generators for cellular assays is that cell-to-cell communication by diffusible factors is disrupted by continuous flow as such factors are washed away. 
Cells also experience a flow-induced shear stress, which may affect cell behaviour.
Diffusion-based designs can overcome these problems and are appropriate for analysing the cell's response to gradients where autocrine or paracrine cell signaling plays an important role.
A typical diffusion based design features rectangular chambers between a source and a sink flow channel.
In devices with such a design \cite{Paliwal:2007aa,Saadi:2007rt}, a technical difficulty is to minimize the unwanted cross-flow between the source and the sink to maintain a stable gradient \cite{Frank:2013kq}.
Different schemes have been proposed to rectify this problem of flow-induced perturbation: membranes  \cite{Abhyankar:2006xy,Kim:2009ys}, hydrogels \cite{Cheng:2007kx} or micropillars \cite{Mosadegh:2007fr,Lee:2012yg} between channels and a diffusion chamber; microfluidic "jets" that inject small amounts of fluid with minimal flow over cells \cite{Keenan:2006aa}; increased flow resistance across the chamber by reducing its height \cite{Mosadegh:2007fr}; embedding cells in a collagen matrix \cite{Saadi:2007rt,Mosadegh:2007fr}; a contact zone to balance the pressures of the source and sink streams \cite{Irimia:2007fk}; a two-layered device with micro-channels in the top layer and a buried diffusion chamber in the bottom layer \cite{Atencia:2012fk}.
\newline\indent 
Another notable solution for this technical challenge was put forward recently by Frank and Tay \cite{Frank:2013kq}.
In their flow-switching device, only one side of the chamber is exposed to the flow channels at a time. 
By alternate switching of the on-chip membrane valves separating the chamber and the source/sink channels, gradients were established while a cross flow in the chamber was eliminated.
Its capability of generating extremely stable spatial gradients was demonstrated with molecules such as lipopolysaccharides (LPS), tumour necrosis factor-$\alpha$ (TNF-$\alpha$) and platelet-derived growth factors (PDGF) along with a fluorescent tracer (FITC-dextran, 40 kDa) \cite{Frank:2013kq}. 
\newline\indent 
The reliable maintenance of gradients by the flow-switching scheme, however, depends on the size of diffusing molecules (i.e., diffusion coefficients), the length of a chamber and the duration between valve switching, as gradients degrade by diffusion in a closed chamber.
 Our calculation suggests that in a chamber of 1 mm length with 60 seconds between flow-switching (as specified in  \cite{Frank:2013kq}), molecular weights need to be larger than $\sim$3.5 kDa (Section \ref{sec:decay}); many biologically active molecules fall below this threshold. 
 Microfluidic gradient devices for unicellular organisms like yeast or slime mould \emph{Dictyostelium} usually have diffusion chambers of less than 1 mm in length (for example \cite{Paliwal:2007aa,Skoge:2010vn}). 
In a shorter chamber of 500 $\mu$m, unless using a shorter period of switching cycles, it may not be possible to maintain gradients of any biochemicals reliably with the flow-switching scheme (Section \ref{sec:decay}).
The interval of flow-switching could be shortened in the flow-switching device for fast-diffusing (low molecular weight) molecules.
Yet, it makes it necessary to reduce the number of parallel gradients that can be controlled in a single chip, compromising one of the very advantages of its design.
\newline\indent 
In this paper we used theoretical models and numerical simulations to aid the design and testing of a device for diffusion-based gradient generation. 
Our design has a channel junction upstream of the diffusion chambers, where the source and sink flows are merged to create a laminar flow away from the cell chambers; this enables precise balancing of flows in the source and sink channels thereby eliminating cross flows in the chambers.
A similar idea was adopted in the device developed by Atencia \emph{et al.} \cite{Atencia:2012fk} and Irimia \emph{et al.} \cite{Irimia:2007fk}; their device also merges the source and sink flows separated by a laminar interface, which in turn generate a gradient in the diffusion chambers downstream.
In their device, however, there is a diffusive mixing between the merged streams, which may be an issue depending on the length of the laminar interface and the flow rates.
To avoid a diffusive mixing, our device has a thin separation wall that splits the merged source/sink flows towards the cell chambers. 
We also adopted the design features reported previously including chaotic mixer channels \cite{Stroock2002aa} with the so-called Dial-A-Wave (DAW) junction \cite{Ferry:2011fj}.
The modular combination of these preexisting microfluidics parts with the unique channel design allows rapid and independent control of biochemical concentration in the source/sink flows.
\newline\indent
We demonstrate the capability of our device using fluorescent tracer dyes and yeast cells harbouring a fluorescent protein reporter gene that are exposed to a gradient of the inducer doxycycline (Dox).
Unlike the flow-switching device, our design allows stable gradient generation irrespective of the molecular mass, as well as a rapid and dynamic control of gradients.

\section{A model of diffusion-based gradient with adsorption} \label{sec:adsorption}
We consider the diffusion of molecules that can be adsorbed on the exposed surface (e.g. PDMS).
The process of adsorption can be described by the law of Langmuir \cite{Tabeling:2005fj}.
The theory of Langmuir assumes that molecules can only be adsorbed on the solid surface that are not yet covered by the molecules.
Let $\phi$ be the fraction of surface already occupied by the molecules.
The flux of molecules $\mathit{J_1}$ being adsorbed by the surface is:
\begin{equation} \label{eq:j1}
\mathit{J_1}=\mathrm{k_1} \mathit{u_f} (1-\phi),
\end{equation}
where $\mathrm{k_1}$ is a rate constant and $\mathit{u_f}$ is the local concentration of the molecule in solution.
The backward flux $\mathit{J_2}$ of the adsorbed molecules from the surface is proportional to $\phi$:
\begin{equation}  \label{eq:j2}
\mathit{J_2}=\mathrm{k_2} \phi,
\end{equation}
where $\mathrm{k_2}$ is another rate constant. 
We take these fluxes $\mathit{J_1}$ and $\mathit{J_2}$ into account to model the diffusion dynamics.

\begin{figure}[p]
    \centering
    \includegraphics[width=0.6\textwidth]{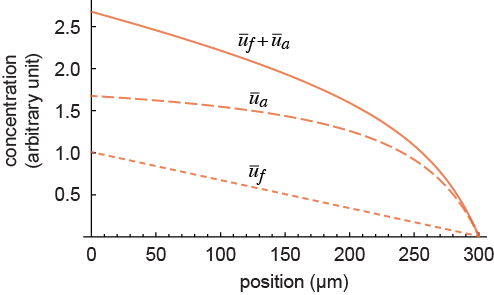}
    \caption{An example of diffusion-based gradient with adsorption at steady state.  $\mathit{\bar{u}_f}$, concentration in solution; $\mathit{\bar{u}_a}$, concentration of adsorbed molecule. $\mathrm{A}=2, \mathrm{c_1}=1, \mathrm{c_2}=0, \mathrm{k_a}=5, \mathrm{L}=300~(\mathrm{\mu m})$.}
    \label{steadystate}
    \end{figure}

The diffusion term can be described by Fick's second law.
The question is how we can combine the law with fluxes $\mathit{J_1}$ and $\mathit{J_2}$.
The height ($\mathrm{H}$) of the cell chambers in our microfluidic device is much smaller than the width ($\mathrm{W}$) and length ($\mathrm{L}$, which is parallel to the direction of diffusion); $\mathrm{H}=4.5 \mu m, \mathrm{W}=100 ~\mu m, \mathrm{L}=300\;\text{to}\;700~\mu m$.
The characteristic diffusive mixing time ($\tau$) of a substance across the height of the chamber is given by $\tau \sim \mathrm{H}^2/\mathrm{D}$, where $\mathrm{D}$ is a diffusion coefficient.
$\mathrm{D}$ is typically on the order of $10^{-6}$ to $10^{-5} cm^2 s^{-1}$ ($10^2$ to $10^3 ~\mu m^2 s^{-1}$) in aqueous solution, for which $\tau \sim 0.02\;$ to $\;0.2 \;s$. 
This is orders of magnitude lower than the diffusion time across the length of $\mathrm{L}$ or $\mathrm{W}$. 
The sides of the chamber (PDMS) may also influence the gradient profile.
But this boundary effect was noticeable only very close to the PDMS because the concentration of molecules averages out along the width of the chamber by diffusion, as observed in the gradient profiles in the experiments described later (for example, see Fig. \ref{FDgrad}c).
We can thus approximate the diffusion with adsorption by a model in one-dimension described by Eqs. ~\ref{eq:ads1}-\ref{eq:ads3}.
\begin{align}
\frac{\partial \mathit{u_f}}{\partial \mathit{t}} &= \mathrm{D} \Delta \mathit{u_f}-\mathit{J_1}+\mathit{J_2}-\mathrm{k_d}  \mathit{u_f},\label{eq:ads1}\\
\frac{\partial \mathit{u_a}}{\partial \mathit{t}} &=\mathit{J_1}-\mathit{J_2}, \label{eq:ads2}\\
\phi &=\frac{\mathit{u_a}}{\mathrm{A}} \label{eq:ads3}
\end{align}
where $\Delta$ is the Laplacian operator, $\mathrm{D}$ is a diffusion coefficient, $\mathrm{k_d}$ is the decay (degradation) rate constant, $\mathit{u_a}$ is the local concentration of the adsorbed (immobile) molecule, and $\mathrm{A}$ is the maximal $\mathit{u_a}$ (that is, the surface concentration at saturation).
At steady state ($\mathit{t}=\infty$), from Eqs. \ref{eq:ads1} and \ref{eq:ads2}:
\begin{align} \label{eq:init}
\mathrm{D} \Delta \mathit{u_f}(\mathit{x},\infty)&-\mathrm{k_d}  \mathit{u_f}=0,\\ 
\mathit{J_1}&=\mathit{J_2}. 
\end{align}
From Eq. \ref{eq:init}, the steady state (concentration) gradient \emph{in solution} is independent of adsorption.
A steady state solution can be obtained using a boundary condition, which corresponds to the concentration at the left edge ($\mathit{x}=0$) and the right edge ($\mathit{x}=\mathrm{L}$) of a diffusion chamber:
\begin{equation}\label{eq:bc}
\mathit{u_f}(0,\mathit{t})=\mathrm{c_1},~\mathit{u_f}(\mathrm{L},\mathit{t})=\mathrm{c_2},
\end{equation}
where $\mathrm{c_1}$ and $\mathrm{c_2}$ are constants.
For fluorescent tracers, we assume $\mathrm{k_d}=0$ as their decay due to photobleaching is negligible compared to the mass transfer by diffusion from the parallel channels. 
Assuming $\mathrm{k_d}=0$ and solving Eqs. \ref{eq:init}-\ref{eq:bc}, the steady state solution is:
\begin{align}
\mathit{\bar{u}_f} &=\mathrm{c_1}+\frac{\mathrm{c_2}-\mathrm{c_1}}{\mathrm{L}}\mathit{x}\,,\label{eq:linear}\\ 
\mathit{\bar{u}_a} &=\frac{\mathrm{A}\, \mathit{\bar{u}_f} }{\frac{\mathrm{k_2}}{\mathrm{k_1}}+\mathit{\bar{u}_f} }\,\\
&=\frac{ \mathrm{A} \,\mathit{\bar{u}_f} }{ \frac{1}{\mathrm{k_a}}+\mathit{\bar{u}_f} } \;,
\end{align}
where $\mathit{\bar{u}_f}=\mathit{u_f}(\mathit{x},\infty), \mathit{\bar{u}_a}=\mathit{u_a}(\mathit{x},\infty)$; $\mathrm{k_a}$ is the adsorption coefficient defined as $\mathrm{k_a}=\frac{\mathrm{k_1}}{\mathrm{k_2}}$ (Fig. \ref{steadystate}).

Therefore, the steady state gradient \emph{in solution} is linear (and independent of adsorption) in the absence of molecular decay.
In the presence of a decay, the gradient is an exponential decay curve.
In the presence of adsorption, however, the effective diffusion would be slower than predicted by the diffusion coefficient $\mathrm{D}$.
For a fluorescent molecule, the total fluorescence is proportional to $\mathit{\bar{u}_f}+\mathit{\bar{u}_a}$, which shows a characteristic concave profile at the steady state (Fig. \ref{steadystate}).
The gradient of the fluorescent tracer Sulforhodamine 101 (SR) is concave as we later show, indicating the adsorption of SR on the surface of the chamber.

\section{Estimation of diffusion coefficients}\label{sec:mw}
Evans \emph{et al.} \cite{Evans:2013kq} provides an equation to calculate diffusion coefficients of small molecules from their molecular weights:
\begin{equation}
\mathit{D}=\frac{\mathrm{k_B} T (\frac{3\alpha}{2}+\frac{1}{1+\alpha})}{6 \pi\,\eta\sqrt[3]{\frac{3 MW}{4 \pi \,\rho_{eff} \mathrm{N_A}}}},\;\alpha=\sqrt[3]{\frac{MW_s}{MW}}, \label{eq:diff}
\end{equation}
where $\eta$ is the dynamic viscosity of solvent (water), $\rho_{eff}$ is the effective density of a small molecule, $\mathrm{k_B}$ is the Boltzmann constant, $\mathrm{N_A}$ is the Avogadro number, $\mathit{T}$ is the absolute temperature; 
$MW_s$ and $MW$ are the molecular weights of the solvent and the molecule for which we want to estimate the value of $\mathit{D}$, respectively (the function plotted in Fig. \ref{coefficient}a).
They showed that calculated coefficients were good estimates of the values determined experimentally \cite{Evans:2013kq}.

To calculate the diffusion coefficient of Sulforhodamine 101 (SR; $MW=606.71 ~\mathrm{Da}$) and Cascade Blue hydrazide (CB; $MW=527.47\;\mathrm{Da}$), we used the following values of constants and parameters: $\mathit{T}=298\,\mathrm{K}, \eta=0.89\times10^{-3} ~\mathrm{kg\,m^{-1}\,s^{-1}}, \rho_{eff}=619~\mathrm{kg\, m^{-3}}, MW_s=18\times 10^{-3} \;\mathrm{kg\,mol^{-1}}$.
The estimated values of diffusion coefficients were $413 \;\mathrm{\mu m^2\, s^{-1}}$ for SR and $437 \;\mathrm{\mu m^2\, s^{-1}}$ for CB.

To estimate the diffusion coefficient of FITC-dextran ($MW=10 ~\mathrm{kDa}$), we used a power law relationship:
\begin{equation}
\mathit{D} =\mathit{a}\, (MW)^\mathit{\,b}\; (\mathrm{cm^2\,s^{-1}})
\end{equation}
with $\mathit{a}=2.71\times10^{-5}, \mathit{b}=-0.37$ \cite{Gribbon:1998qf}.
The calculated value of the diffusion coefficient is $8.97\times10^{-7}\;\mathrm{cm^2\,s^{-1}}=89.7 \;\mathrm{\mu m^2\, s^{-1}}$.

\section{Decay of gradients in a closed chamber}\label{sec:decay}
\begin{figure}[h]
    \centering
    \includegraphics[width=0.6\columnwidth]{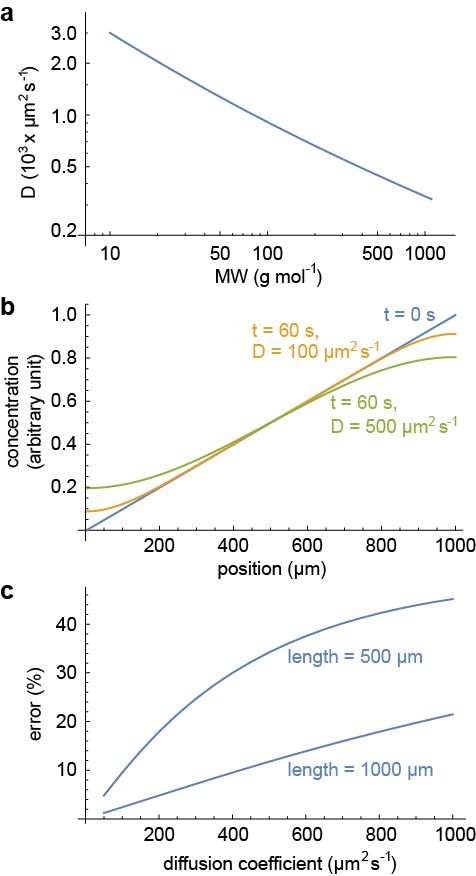}
    \caption{Estimation of diffusion coefficients and error \% introduced to the gradient after isolating a chamber from source/sink channels. (A) Diffusion coefficient $\mathrm{D}$ is plotted as a function of molecular weight (MW). (B) Gradient decay due to diffusion in a confined chamber of length = 1000 $\mu$m. (C) \% error introduced after 60 s in chambers of length = 500 $\mu$m or 1000 $\mu$m.}
    \label{coefficient}
    \end{figure}

The flow-switching scheme for gradient generation and maintenance described by Frank and Tay \cite{Frank:2013kq} operates by alternate opening and closing of the on-chip membrane valves.
The valves are open for 1 s each every 2 min, therefore, chambers are isolated from the source and sink channels for $\sim$1 min.
We calculated how much a gradient diverges from a steady state during this period as they decay by diffusion in a confined environment.
This was done by solving the diffusion equation $\frac{\partial \mathit{u}}{\partial \mathit{t}} = \mathrm{D} \Delta \mathit{u}$ with a boundary condition:
\begin{equation}
\frac{\partial u}{\partial x}\big\vert_{x=0}=\frac{\partial u}{\partial x}\big\vert_{x=\mathrm{L}}=0,
\end{equation}
and an initial condition, which is a linear gradient:
\begin{equation}
u(x,0)=\frac{\mathrm{C}\, x}{\mathrm{L}},
\end{equation}
where C is a constant. In the flow-switching device, $\mathrm{L}=1$ mm. 
Fig. \ref{coefficient}b shows the plot of the solution for $\mathrm{D}=100$ and $500$ $\mathrm{\mu m^2\, s^{-1}}$ (with $\mathrm{C}=1$).
\% error is the fraction of the area diverged from the linear gradient after 60 s, which is plotted against diffusion coefficient in Fig. \ref{coefficient}c.
For a chamber with L = 1000 $\mu$m, $\mathrm{D}$ = 208.4 $\mathrm{\mu m^2\, s^{-1}}$ results in 5 \% error.
From Eq. \ref{eq:diff}, this corresponds to the molecule weight of $\sim$3.5 kDa.

 \section{Device design}

\begin{figure}[t]
     \centering
       \includegraphics[width=0.9\columnwidth]{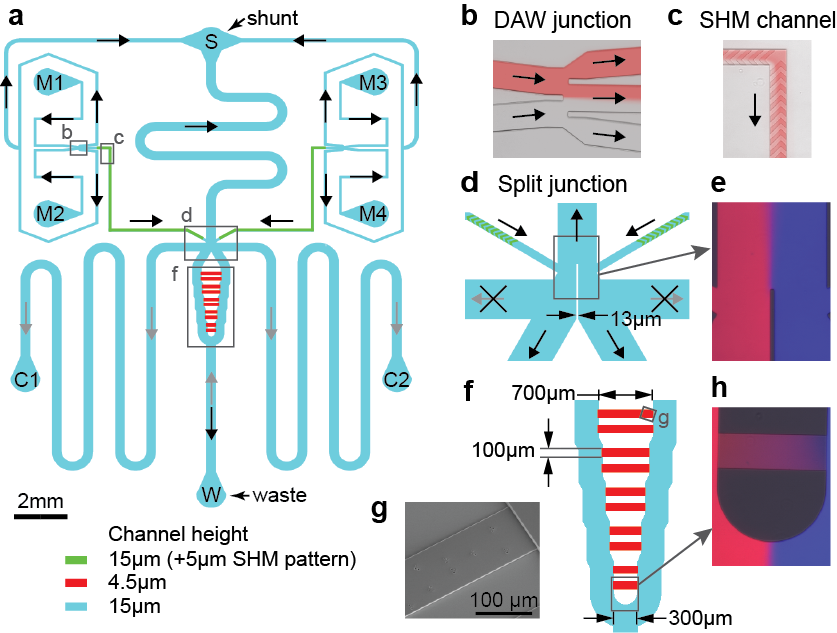}
     \caption{Design of microfluidics. (a) Overview of the microfluidics. Main channels (light blue), chaotic/staggered herringbone mixer (SHM) channels (green) and cell traps (red) are shown. Black arrows indicate the flow direction for gradient generation and cell culture; grey arrows indicate the flow direction during cell loading. Flows towards ports C1 and C2 are stopped after cell loading. Grey rectangles indicate the positions of panels b, c, d and f. (b) Pseudo-color image of medium flow (Sulforhodamine 101 (SR) in red) at a DAW junction. (c) Medium flow in SHM channel (SR in red). (d) Overview of the split junction. Laminar flow (SR in red; Cascade Blue (CB) in blue) at the junction is shown in panel e. (f) Overview of the diffusion chambers (red) between a pair of parallel channels. The chambers in the alternative positions have pillars as shown in panel g (SEM image of a silicon wafer template) for trapping cells. (h) Gradients of two fluorescent tracer dyes (SR, red; CB, blue) in one of the cell traps indicated in panel f.}
    \label{ChipDesign}
    \end{figure}

The design is based on the devices developed by Paliwal et al. \cite{Paliwal:2007aa} and Ferry et al. \cite{Ferry:2011fj}.
 The microfluidics device consists of main channels, a pair of mixer modules and rectangular diffusion/cell chambers (Fig. \ref{ChipDesign}). 
Flows in the chip are controlled by hydrostatic pressure.
 A single unit of the controller holds two media reservoirs that are connected to a pair of inlet ports (M1-M2 or M3-M4 in Fig. \ref{ChipDesign}a). 
 Media inputs from these ports flow through a \textquotedblleft Dial-A-Wave\textquotedblright \;(DAW) junction \cite{Ferry:2011fj} (Fig. \ref{ChipDesign}a and b).
 At the DAW junction two media inputs are combined and mixed while passing through a staggered herringbone mixer (SHM) channel \cite{Stroock2002aa} (Fig. \ref{ChipDesign}a (highlighted in green) and c).
 The mixing ratio is controlled precisely by the hydrostatic pressure controller adopted from Ferry et al. \cite{Ferry:2011fj} (Fig. S7).
  The mixer modules are positioned symmetrically, which feed two mixed media to a channel junction in the middle of the device (which we call the \textquotedblleft split junction\textquotedblright; Fig. \ref{ChipDesign}d).
 See the Supplementary movie SplitJunction.avi, which is a time-lapse image sequence of SR fluorescence at the split junction; the fluorescent dye was supplied as a triangle wave (5 min period, 2 cycles).

Once the mixed media reaches the split junction, the media flows towards the shunt port (S) or down a pair of parallel channels leading to the diffusion chambers (highlighted in red in Fig. \ref{ChipDesign}a and f).
The split junction has a separator wall that eliminates a diffusive mixing of the two media flows (Fig. \ref{ChipDesign}d and e).
Without this separator wall, significant diffusive mixing at the laminar interface was observed (Supplementary material, Fig. S1c, d).
The separation wall at the split junction may be extended for an additional length (perhaps 10 to 100 $\mu$m, which should be determined experimentally).
This modification allows counter flows towards the shunt port to eliminate the possibility of minor diffusive mixing between the left and right channels at the end of the separation wall.

The split junction is also connected to cell outlet ports through serpentine channels (C1 and C2, Fig. \ref{ChipDesign}a).
Once cells are loaded in the chip, C1 and C2 ports are closed by small binder clamps on the Tygon tubes just outside of the chip to stop the flows through these ports.
The parallel channels act as a source and a sink for diffusion through the rectangular chambers of different lengths.
 The parallel flows merge downstream of the chambers towards the waste port (W).
 
 At the split junction a laminar flow is created towards the shunt port (Fig. \ref{ChipDesign}e).
 Together with a laminar flow at another junction downstream of the chambers towards the waste port (Fig. \ref{ChipDesign}h), the laminar flow interface serves as an indicator of the balance of the right and left media flows towards the diffusion chambers.
The position of the laminar flow interface can be adjusted by fine-tuning the height of the media reservoirs on the hydrostatic pressure controllers thereby equalising the left-right pressure balance.
This eliminates the cross-flow across the chambers and enables an establishment of stable diffusion-based chemical gradients.

\section{Materials and Methods}\label{sec:mm}
  \subsection{Microfluidic device fabrication}
The microfluidic devices were fabricated using conventional soft-lithography methods with poly(dimethylsiloxane) (PDMS) \cite{Duffy:1998rt}.
Silicon wafers (10 cm diameter) patterned with SU8 photoresist were custom-ordered from Kelvin Nanotechnology Ltd. (Glasgow, UK).
AutoCAD (Autodesk) was used to draw the device design.
CAD files are available on request.
Device chips were made by pouring PDMS (Sylgard 184, Dow Corning; 10:1 mix with curing agent) over the mold.
PDMS was then degassed, cured at \SI{70}{\celsius} for 3 hours, peeled from the mold and cut into individual chips. 
Holes for media inlets and outlets were punched through the chip using a sharpened Harris UNI-CORE\texttrademark ~biopsy punch (0.5 mm).
Dust was removed from PDMS chips and coverslips by ionizing air (nitrogen) gun (Top Gun, Simco) before assembly.
PDMS chips and coverslips were bonded together after treatment with a plasma cleaner (Zepto ONE, Diener Electronic; HF generator 50 \% \,($\sim$15 W), 30 seconds).
Assembled chips were then cured further at \SI{60}{\celsius} ~overnight. 
Microbore tubing (ID/OD 0.02"/0.06"; Cole-Parmer, Tygon S-54-L) was used to connect a PDMS chip and media reservoir syringes.
Hydrostatic pressure actuators are fabricated according to Ferry et al. \cite{Ferry:2011fj}, which is controlled by a Raspberry Pi computer (Raspberry Pi 2 Model B); \url{https://www.raspberrypi.org}) and driven by a Java code modified from \cite{Ferry:2011fj} (\url{http://biodynamics.ucsd.edu/dialawave/}).
The Java source code is available upon request. 
Setup and operation protocol of the platform is described in detail in Supplementary material Section 5.

\subsection{Yeast strains and media}
Two diploid strains of yeast \emph{S. cerevisiae}, MM169 and MM201, were used in this study, each of which harbours a fluorescent protein reporter gene (destabilised version of yVenus \cite{Giuraniuc:2013qf}, an enhanced YFP).
 MM169 also has a gene encoding the transcriptional activator rtTA of the Tet-On system \cite{Giuraniuc:2013qf} while MM201 has a tandem array of genes encoding rtTA and TetR-Ssn6 (a synthetic transcriptional repressor) that constitutes the activator/repressor dual system \cite{Belli:1998qf}.
See Supplementary material Fig. S6 for the illustration of these gene constructs (Fig. S6a for MM169; Fig. S6c for MM201), which were all integrated in the genome using the method described previously \cite{Giuraniuc:2013qf,MacPherson:2016uq}.
Details of the construction of these strains will be described elsewhere.
For microfluidics experiments, yeast cells were first grown overnight at \SI{30}{\celsius} in 10 ml synthetic complete media with raffinose (SMR), the composition of which is identical to the synthetic complete (SC) media \cite{cslmanual} except glucose is replaced with raffinose (2 \%; Formedium, cat. no. RAF01). 
On the following day the SMR cell culture was spun down at 3,000 g for 10 min and resuspended in 2 ml synthetic complete media with maltose (SMM; same as SC with maltose (2 \%; BD Difco cat. no. 216830) replacing glucose) supplemented with BSA (Sigma A7906; final concentration at 0.01 \%).
Cells were then injected into a microfluidic device by 20 ml syringe through 10 cm of Tygon tube connected to the waste port.
The cell loading procedure is detailed in Supplementary material Section 5.3.
Once cells were loaded in the device, SMM was used for cell growth.
Doxycycline was added to one of the reservoir media syringes to the final concentration of 2 $\mu$g/ml.
For shake flask experiments (Supplementary material Fig. S6) yeast strains were cultured in 10 ml of SMR at \SI{30}{\celsius} overnight.  
 These precultures were then diluted to OD$_{600} = 0.4$ in 10 ml SC, SMR or SMM and continued to be grown at \SI{30}{\celsius}.
Each cell culture was split after 2 hours and either treated with doxycycline (Dox) (at the final concentration as indicated in Fig. S6) or left untreated. 
After further 7 hours culture at \SI{30}{\celsius}, images were captured by fluorescence microscopy using the same imaging system for microfluidics experiments.

 \subsection{Microscopy and data analysis}
Fluorescent tracer molecules were used at the following final concentrations in SC or SMR media: Cascade Blue hydrazide (CB; Trilithium salt, Thermo-Fisher) at $20 ~\mu$M, Sulforhodamine 101 (SR; Sigma-Aldrich S7635) at $10 ~\mu$g/ml and Fluorescein-dextran (FD; molecular weight 10,000, Sigma-Aldrich FD10S) at 1 mg/ml.
For microfluidic experiments with yeast cells, CB was added to SMM.
For microscopy we used a Zeiss Axio Observer Z1 microscope equipped with an incubation chamber, CCD camera (Coolsnap HQ2, Photometrics) and Axiovision software (v4.8.2 SP3 (08-2013)).
Either 20x (Ph2) or 40x (Ph3) objective was used for imaging. 
Fluorescence images were captured using Zeiss 62 HE BFP/GFP/HcRed filter set (excitation filter: BP370/40 for CB; BP474/28 for FD and Venus; BP585/35 for SR).  
 
To quantify the gradients of fluorescent tracers, a rectangular region (width = 100 $\mu$m) in the captured images corresponding to a single diffusion chamber was cropped using ImageJ (version 1.50g). 
Average (total) pixel intensities were calculated along each single pixel line parallel to the main flow channels (orthogonal to the direction of diffusion).
Fluorescence intensities in the main flow channels are represented by the intensity of a single pixel line parallel to the flows, approximately 20 $\mu$m away from the diffusion chamber's edge.
Theoretical calculations of channel flows and diffusion/adsorption models were performed using Mathematica software (version 10.4, Wolfram Research).

\section{Results and Discussion}

\subsection{Computer-controlled gradient of small molecules}
\begin{figure}[t]
    \centering
    \includegraphics[width=0.7\textwidth]{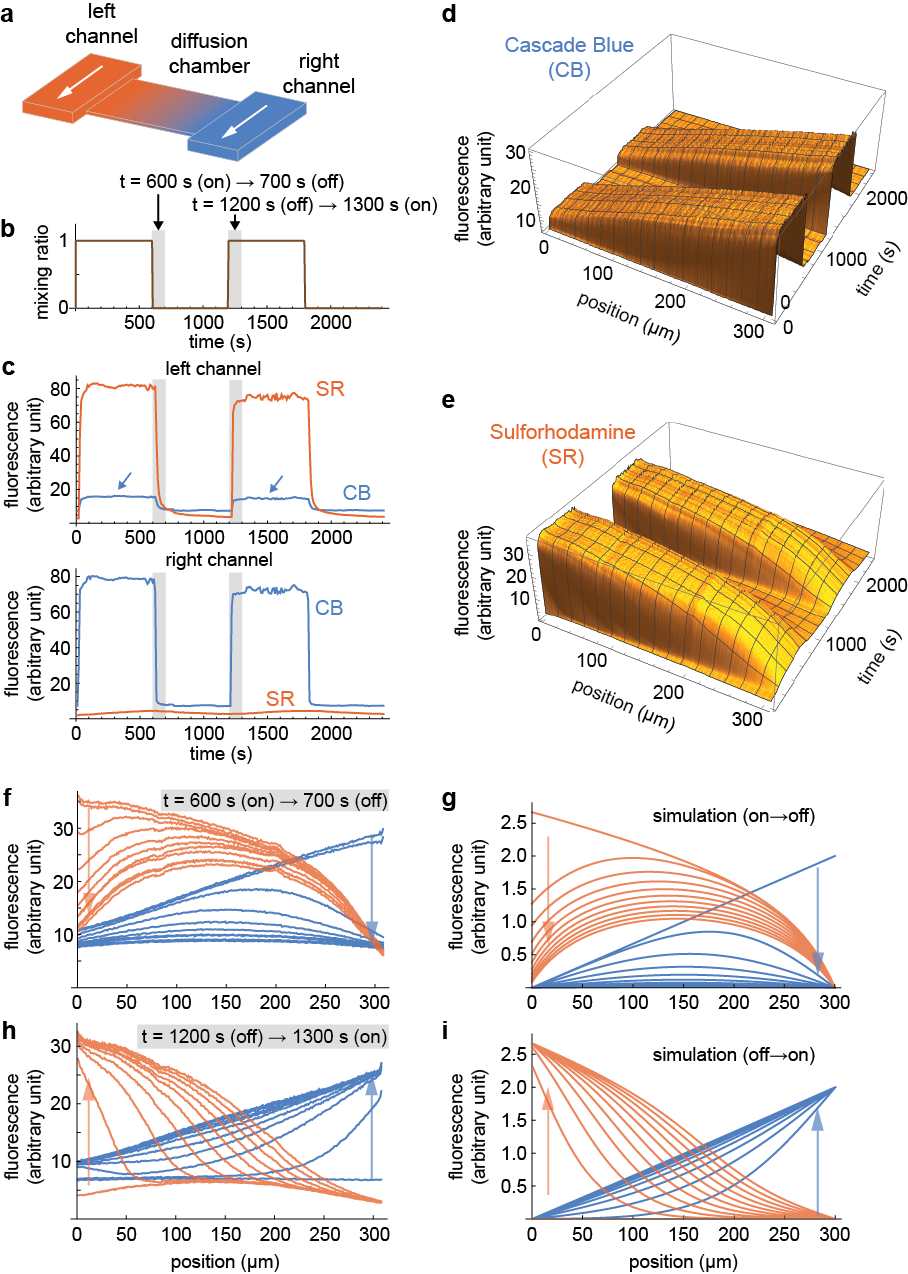}
    \caption{Generation and control of gradients in a diffusion chamber. (a) A schematic of gradient generation by diffusion. (b, c) Fluorescent dyes were supplied in square waves (b) for both channels and their fluorescence was measured in the left (c, upper panel) and right channels (c, lower panel) over time. Arrows in c, upper panel, indicate the fluorescence SR signal bleed-through in the detected CB fluorescence (see Supplementary material Section 3). (d, e) The fluorescence intensity measured along the chamber over time. (f, h) Fluorescence across the chamber was measured every 10 s over 100 s periods, which are highlighted as grey bars in panels b and c. (g, i) Numerical simulations of fluorescence intensity profiles (10 s intervals, over 100 s period) corresponding to the measured ones in panels f and h, respectively. The direction of the change in gradient profiles are indicated by red or blue arrows in panels f to i. Parameters: $\mathrm{c_1}=1$, $\mathrm{c_2}=0$, $\mathrm{k_a}=5$, $\mathrm{A}=2$ (for SR) or $\mathrm{c_1}=0$, $\mathrm{c_2}=2$ (for CB), with diffusion coefficients as mentioned in the text.}
    \label{IndepGrad}
    \end{figure}

First, to test the microfluidic platform, we used fluorescent dyes Sulforhodamine 101 (SR; M.W. = 606.7 Da) and Cascade Blue hydrazide (CB; M.W. = 527.5 Da).
We note that SR and CB are relatively small molecules with similar molecular masses.
Thus their diffusion coefficients are also similar (estimated values are $413 \;\mathrm{\mu m^2\, s^{-1}}$ for SR and $437 \;\mathrm{\mu m^2\, s^{-1}}$ for CB; see Section \ref{sec:mw}).
SR and CB flowed through the left (SR) and right (CB) channels to generate gradients (Fig. \ref{IndepGrad}a).

The fluorescent dyes were supplied as square waves (Fig. \ref{IndepGrad}b and c), which were controlled independently. 
Fluorescence of CB and SR was quantified by fluorescence microscopy in one of the chambers (300 $\mu$m in length).
Note that a slight increase in CB fluorescence in the left channels (arrows in Fig. \ref{IndepGrad}c, upper panel) is due to a bleed-through of SR signal (see Supplementary material Section 3). 
Fig. \ref{IndepGrad}d and e show the fluorescence along the chamber over time.
The gradient of fluorescence is linear for CB while the one for SR is concave.
Similar gradients were detected in the other chambers of different lengths (data not shown).
If the mass transfer of a chemical occurs only by passive diffusion then it is predicted that at the steady state a linear gradient is established in a rectangular chamber between source and sink channels \cite{Frank:2013kq}, as observed with CB (Fig. \ref{IndepGrad}d).
The most plausible explanation of the concave profile of SR fluorescence is the adsorption of SR on the solid surface of the chamber (Section \ref{sec:adsorption}).
The detected SR fluorescence is the sum of the signals from the free molecules in solution and those adsorbed on the surface.

\subsection{Diffusion in the presence of adsorption}
The diffusion-adsorption in a chamber can be modelled by Fick's law of diffusion and the law of Langmuir that describes the equilibrium between the molecules in solution and the adsorbing surface \cite{Tabeling:2005fj} (see Section \ref{sec:adsorption}).
This simple model recapitulates the observed fluorescence profiles of SR rather well. 
Fig. \ref{IndepGrad}f and h show the fluorescence of SR (red) and CB (blue) across the chamber over the 100 s period (10 s interval) after the tracer inputs were switched off (Fig. \ref{IndepGrad}f) or on (Fig. \ref{IndepGrad}h).
The observed fluorescence profiles over time are qualitatively similar to the simulation results (Fig. \ref{IndepGrad}g and i).
Although SR and CB have similar diffusion coefficients, it took much longer to establish a steady state gradient for SR than CB.
Thus, when adsorption occurs, the effective diffusion rate of a molecule can be smaller than predicted by assuming a free diffusion. 
In this particular experiment with a period of 1200 s, the SR gradient did not reach the steady state profile while the CB gradient did.
In most cases, we have no prior knowledge about the adsorption of the molecules of interest at the solid-liquid interface.
Therefore, a fluorescent tracer of a similar molecular weight is not necessarily a good proxy of the molecule to be analysed by gradient generation.
If a biochemical of interest interacts with the substrate such as PDMS (for example, retinoic acid \cite{Futrega:2016qy}) a careful analysis of gradient dynamics may be necessary to interpret experimental results properly.
Even so, the steady-state gradient of molecules \emph{in solution} is predicted to be linear and independent of adsorption (Fig. \ref{steadystate}).

 \subsection{Generation of a gradient of larger molecules}
\begin{figure}[t]
    \centering
    \includegraphics[width=0.9\textwidth]{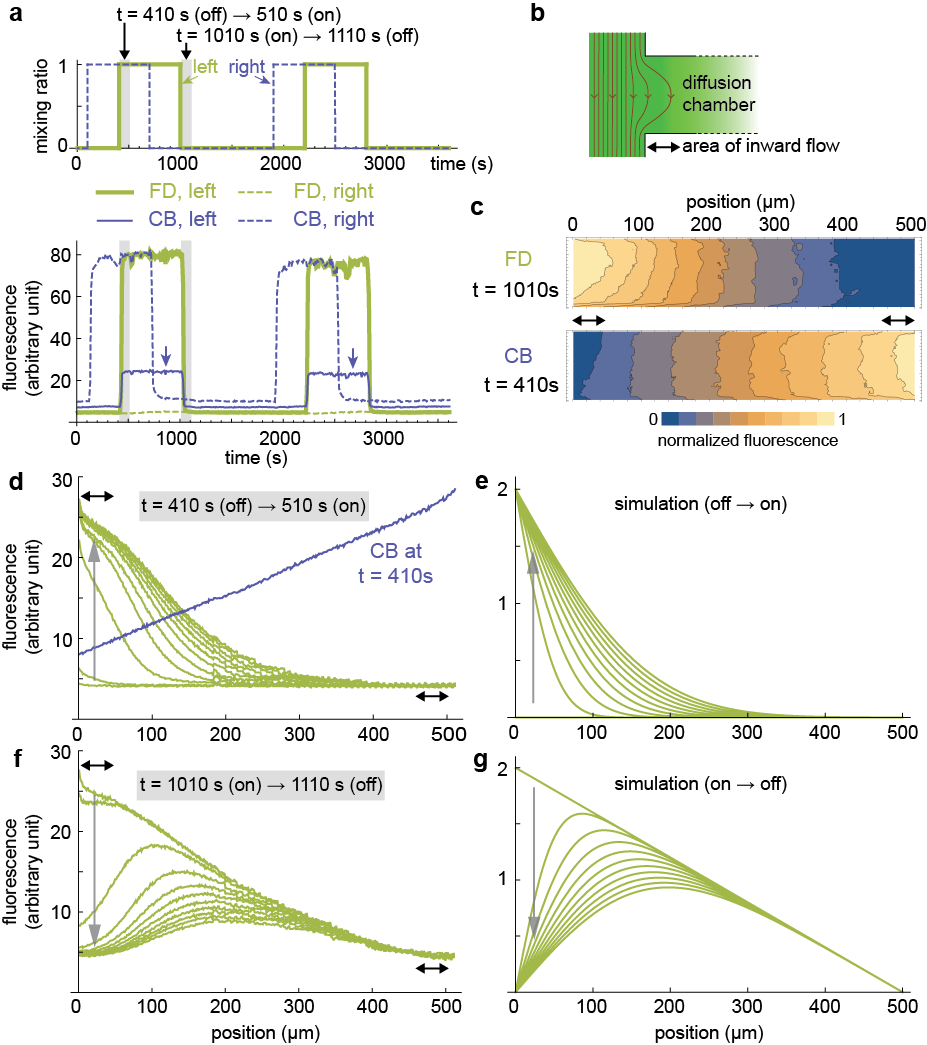}
    \caption{Generation of gradient of FITC-dextran. (a) Fluorescent dyes were supplied in square waves (upper panel; FD in the left and CB in the right channel) and fluorescence in the left and right channels were measured (lower panel). Arrows in the lower panel indicate the bleed-through of FD signal detected as the CB fluorescence. (b) Schematic illustration of inward flow trajectories at the edge of the chamber. The area of inward flow is indicated by a double headed arrow. (c) Contour plots of fluorescence level across the chamber (100~$\mu$m x 500~$\mu$m). The advection dominated areas are indicated by black arrows. (d, f) Fluorescence across the chamber was measured every 10 s over 100 s periods highlighted in grey bars in panel a, and plotted. In panel d, linear CB gradient at t = 400 s is also shown. (e, g) Predicted gradient by diffusion over 100 s period (10 s intervals) after the input is switched on (e) or off (g). The direction of the change in gradient profiles are indicated by grey arrows in panes d to g. Diffusion coefficient = $89.7 \;\mathrm{\mu m^2\, s^{-1}}$ was used for the calculation.}
    \label{FDgrad}
    \end{figure}

 Next, we tested a gradient generation using FITC-dextran (FD; M.W. = 10 kDa) and CB as a reference. 
 The estimated diffusion coefficient of FD is $89.7 \;\mathrm{\mu m^2\, s^{-1}}$ (Section \ref{sec:mw}).
 FD and CB were supplied in square waves (300 s out of sync) to the left (FD) and right (CB) channels (Fig. \ref{FDgrad}a, upper panel).
 Note that we detected a significant bleed-through of FD signal in the CB fluorescence (Fig. \ref{FDgrad}a, arrows in lower panel; see Supplementary material), which hampers an accurate measurement of CB fluorescence in the presence of FD.
 Fluorescence was measured in a chamber of length = 500 $\mu$m.

 Fig. \ref{FDgrad}d and f show the fluorescence profile across the chamber for every 10 s over the 100 s periods as indicated (grey bars in Fig. \ref{FDgrad}a).
 Fig. \ref{FDgrad}e and g are the corresponding gradients predicted by a passive diffusion without adsorption.
 The CB gradient in the absence of FD (at t = 410 s) was linear (Fig. \ref{FDgrad}d).
 By contrast, FD gradients approached a sigmoidal shape after the FD input to the left channel was switched on (Fig. \ref{FDgrad}d, green curves).
The fluorescence intensity profile in the chamber after the input was switched off (Fig. \ref{FDgrad}f) is similar to the one theoretically predicted (Fig. \ref{FDgrad}g) except the flatter regions close to the edges of the chamber (black arrows in Fig. \ref{FDgrad}d, f).

The flatter regions in the concentration profile are also evident in contour plots across the chamber (Fig.  \ref{FDgrad}c, indicated by arrows).
These regions were due to inward flow trajectories near the edges (illustrated in Fig. \ref{FDgrad}b), which creates advection dominated areas.
Such advection-dominated areas near the edges of diffusion chambers were reported previously with FD of M.W. = 40 kDa \cite{Frank:2013kq}.
Apart from the difference due to the effect of advection, the predicted gradient profiles (Fig. \ref{FDgrad}e, g) reproduced observed ones (Fig. \ref{FDgrad}d, f) quite well.
The fluorescence profile of CB is little affected by the inward flow, perhaps because its diffusion is fast enough to counteract the effect of the advection.

 \subsection{Dynamic gradient generation}
Next, we tested the capability of the microfluidic platform to generate a dynamic gradient.
CB media supplied to the left and right channels were controlled independently, so the gradient of varying magnitude, steepness and direction can be generated according to a pre-programmed scheme.
The chemical concentrations in the parallel channels (i.e., boundary conditions) can be changed quickly, which was demonstrated in the response to square wave input signals (see Supplementary material Section 2, Fig. S2 and the movie SquareWaveGrad.avi, which shows the dynamic gradient in Fig. S2).
Fig. \ref{DynGrad} shows an example of a dynamic gradient of CB in a 300 $\mu$m chamber in response to trapezoid waves of inputs.
The Supplementary movie (TrapezoidWaveGrad.avi) provided online shows the time-lapse image sequence of the gradient shown in this figure (1 sec interval, 301 frames = 300 sec; 20 fps).
Pre-programmed schemes are not restricted to square or trapezoid waves. 
In fact, any arbitrary function can be adopted to control the input concentration (mixing ratio) of the substances of interest\cite{Ferry:2011fj} to change the magnitude and steepness of the gradient.

\subsection{Gene expression in budding yeast cells in response to an inducer gradient}
We designed the microfluidic device in order to study the behaviour of engineered budding yeast (\emph{S. cerevisiae}) cells in chemical gradients.
To test the microfluidic device for bioassays, we cultured yeast cells in the diffusion chambers.
The yeast strains harbour a reporter gene encoding a fluorescent protein (Venus, a yellow fluorescent protein), which is expressed in response to a chemical inducer Dox.
In the media used for microfluidic experiments (SMM, see Section \ref{sec:mm} and Supplementary material Section 4), the doubling time of cell growth is 5.4 hours.
As cells grow they gradually occupy the entire chamber and those pushed out of the chamber are washed away by the flows in the main channels.

\indent When cells were exposed to Dox uniformly, all cells expressed the reporter fluorescent protein Venus (Fig. S5, at t = -225 min).
After removing Dox from the chamber, the fluorescence diminished rapidly (Fig. S4, compare t = -225 min and 0 min) indicating the reporter gene expression was switched off.
When Dox gradient was generated, expression of Venus was observed in a population of cells close to the left channel, i.e. the source of the Dox inducer (Fig. \ref{cell}a).
As the source concentration of Dox increased, the region of expression expanded (Fig. \ref{cell}a; the positions of the expression boundary with the initial gradient with 0.2-0 $\mu$g/ml of Dox are indicated by red arrowheads).
Images throughout the whole experiment are shown in Fig. S4.
Dox was supplied with CB as a tracer, which formed a gradient across the chamber with cells (Fig. \ref{cell}b).
Together, these results demonstrate a generation of Dox gradient, which induced a reporter gene expression in yeast cells in a chamber in a concentration-dependent manner.

When source (left) channel concentration of Dox was increased to 1 $\mu$g/ml, some cells close to the sink (right) channel started to express the reporter gene (Fig. \ref{cell}a, at 2670 min, red bracket). 
This is most likely due to a fraction of Dox diffusing out of the chambers upstream.

\section{Conclusions}
\begin{figure}[t]
    \centering
    \includegraphics[width=0.5\columnwidth]{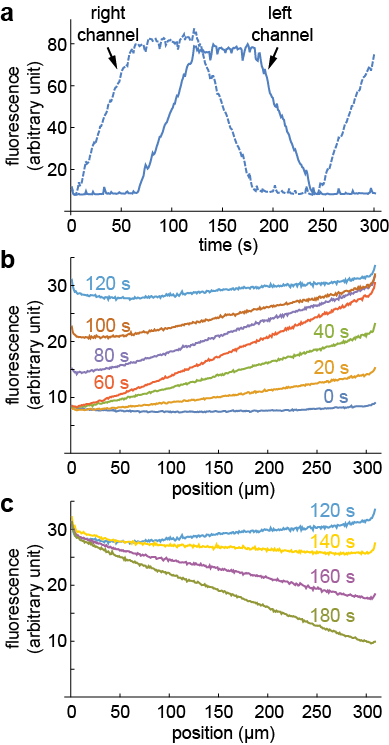}
    \caption{Generation of a dynamic gradient. (a) CB fluorescence in the left and right channels over time. The top and bottom of the  trapezoid fluorescence signals correspond to mixing ratio = 1 and 0, respectively. (b, c) The profile of CB fluorescence across one of the 300 $\mu$m chambers at the indicated time points.}
    \label{DynGrad}
    \end{figure}

We have demonstrated the capability of our microfluidic device that can generate highly dynamic diffusion-based gradients. 
A unique design of the \textquotedblleft split junction\textquotedblright ~in this device allows balancing of the flows in the source and sink channels, thereby eliminating unwanted cross flows in the diffusion chambers that may disrupt gradient profiles and cell-to-cell signaling.
Together with the DAW mixer modules, the split junction enables independent, flexible and rapid change of chemical concentrations in the source and sink channels.
This dynamic feature, as illustrated in Fig. \ref{DynGrad}, is a distinct advantage of our device design over those previously reported.
Unlike the flow-switching device \cite{Frank:2013kq}, our device can generate gradients of fast-diffusing (low molecular weight) molecules.

Our device design, however, has some limitations that include: 1) the steady-state gradient is limited to linear profiles; 2) precise balancing of left and right source/sink flows requires the visualization of the laminar interfaces by fluorescent tracers. 
This prototype device may also benefit from further design modification to maximise its performance.
For example, on-chip pneumatic valves instead of manual clamping of the tube would also simplify the cell-loading process but the advantages need to be weighted carefully against adding an extra layer of complexity.

\begin{figure}[p]
    \centering
    \includegraphics[width=0.4\columnwidth]{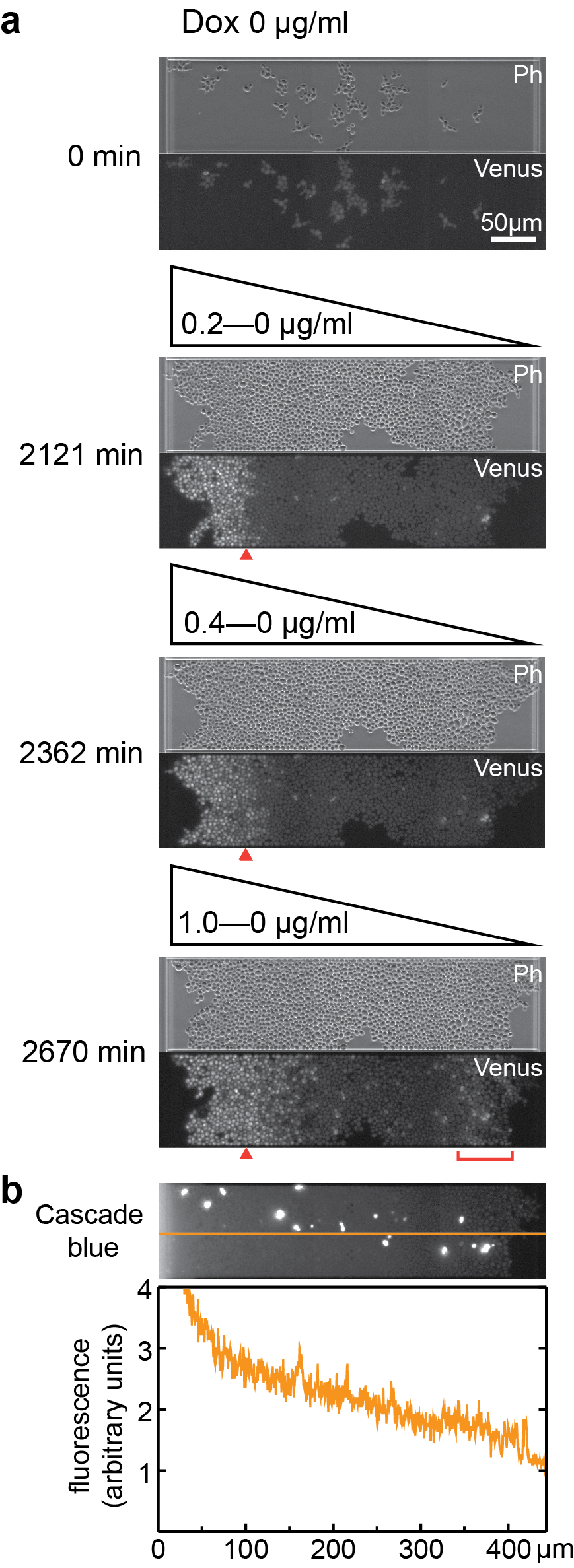}
    \caption{Fluorescent reporter protein expression in yeast in response to the gradient of an inducer. The chamber length is 400 $\mu$m. (a) Phase contrast (Ph) and Venus fluorescence images are shown for the indicated time points. Doxycycline (Dox) concentration are also indicated for each image pair. Red arrowheads indicate the position of approximate boundary of expression at T = 2121 min. The low expression of Venus in the region indicated by a red bracket is likely due to Dox diffused from upstream cell traps. (b) Profile of CB fluorescence  on the image (along the yellow line, above) is plotted (bottom). The image was taken at 2670 min. Bright fluorescence indicates dead cells.}
    \label{cell}
    \end{figure}

We also showed that the adsorption of a molecule on the solid surface may greatly affect its gradient profile. 
Thus, a fluorescent tracer of a similar molecular weight is not always a good proxy of the molecule to be analysed by gradient generation.
As we only used a simple inducible reporter gene expression in yeast (Fig.  \ref{cell}), we have not yet exploited the full capacity of the device for bioassays.
We expect this microfluidic device to be a versatile prototype applicable to a broad range of biomedical investigations of cellular response to gradients such as morphogenesis, chemotaxis or axon guidance.

\begin{acknowledgement}

We thank Marco Thiel and Alessandro De Moura for helpful discussions and advice, Stefan Hoppler and Mamen Romano for critical reading of the manuscript, James Hislop for his help with the plasma cleaner, Alex Brand for the microscopy imaging system and Alistair Robertson for fabricating the hydrostatic flow controllers. We also thank Diane Massie and Yvonne Turnbull for technical assistance. This work was supported by Scottish Universities Life Sciences Alliance (SULSA) and the University of Aberdeen.

\end{acknowledgement}

\newpage

\renewcommand\thefigure{S\arabic{figure}}
\setcounter{figure}{0}
\setcounter{section}{0}

\renewcommand\thetable{S\arabic{table}}

\section*{\huge{Supplementary Material}}

 \section{Design of split junction}
 \subsection{Role of the separation wall}
\begin{figure}[h]
    \centering
    \includegraphics[width=0.7\columnwidth]{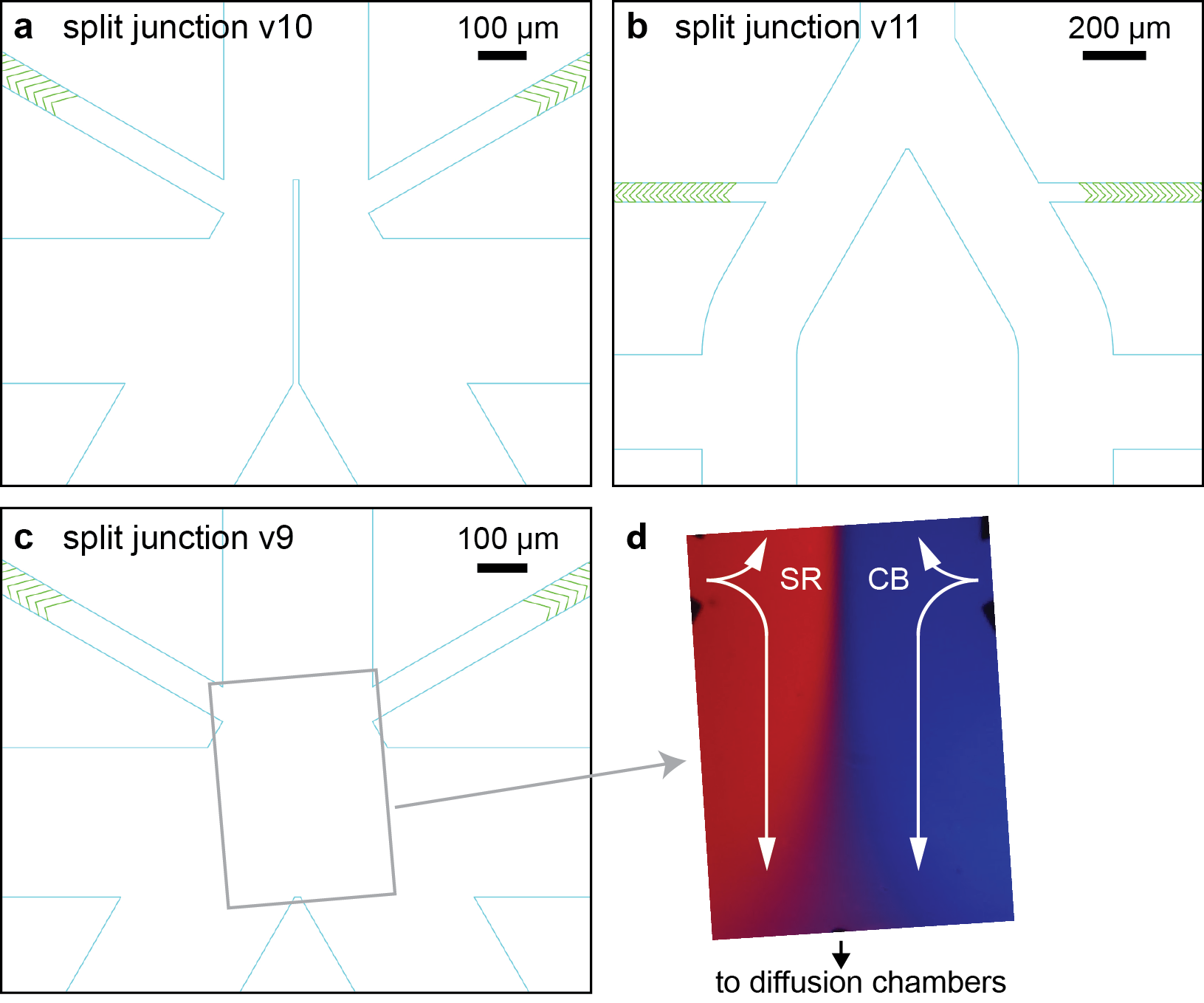}
    \caption{Different designs of the split junction. (a, b, c) show CAD image of each design. The design v9 has no separation wall, which resulted in extensive diffusive mixing at the laminar interface as shown in panel d (purple area).}
  \label{junction}
    \end{figure}

We tested three different designs of the split junction; other parts of the device are identical). 
The overall performances of the designs v10 and v11 (Fig. \ref{junction}a, b) were similar, but the laminar flow at the junction was somewhat easier for observation to balance the flows with the design v10, which has a separation wall (Fig. \ref{junction}a).
The separation wall is necessary to prevent the diffusive mixing at the laminar interface, as evident in the design v9 (Fig. \ref{junction}c, d). 
We therefore used devices with the split junction v10 most of the time.

\subsection{Time-lapse movie of flows at the split junction}
The movie SplitJunction.avi (Supplementary content online) shows time-lapse images of SR fluorescence at the split junction. 
The fluorescent dye was supplied as a triangle wave (5 min period $\times$ 2 cycles).
Note that the laminar flow interface was not affected as the concentration of SR was altered.

  \begin{figure}[h]
    \centering
    \includegraphics[width=0.6\columnwidth]{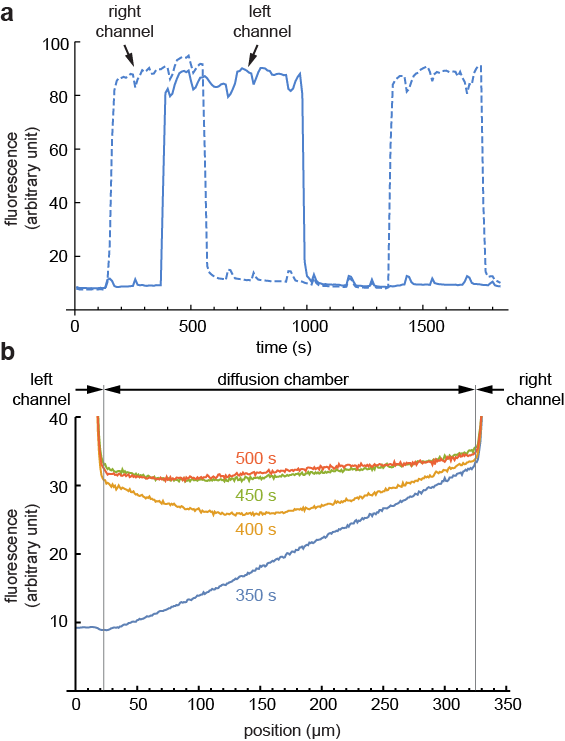}
    \caption{Gradient dynamics in response to square wave signals. (a) Fluorescence of CB in the left and right channels over time. (b) Fluorescence profile across the chamber at the indicated time points.}
  \label{sqwave}
    \end{figure}

\section{Speed of media change in parallel channels}

In our microfluidic chip, media composition can be changed quickly because the media composition (i.e., mixing ratio) is controlled by the DAW junction and SHM channels on chip.
The time required to replace the media flowing through the parallel channels is thus dictated by the flow rate in the channels. 
Typically it takes around 30 seconds to completely replace the media in parallel channels after the execution of a computer program to change the mixing ratio of input media.
The concentration of chemical compounds at the ends of diffusion chambers is therefore controlled precisely and quickly.
This is illustrated by a gradient profile in a diffusion chamber (300 $\mu$m) in response to square wave signals of CB (Fig. \ref{sqwave}).
The movie (SquareWaveGrad.avi) provided as Supplementary content online shows the time-lapse image sequence of
the gradient shown in this figure (10 sec interval, 183 frames = 30 min 20 sec; 20 fps).

\section{Cross-bleeding of fluorescence signals}
\begin{figure}[h]
    \centering
    \includegraphics[width=0.6\columnwidth]{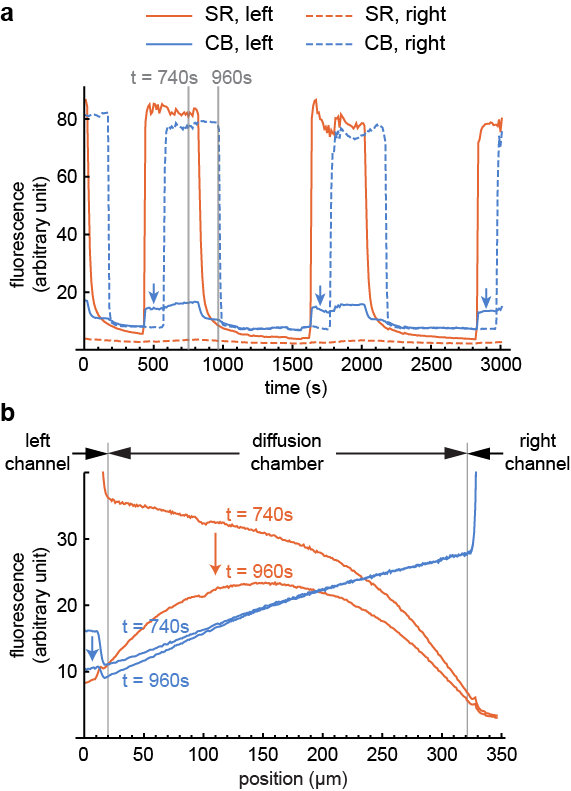}
    \caption{Bleeding of SR fluorescence during CB signal detection. (A) Fluorescence of SR (red) and CB (blue) in the left and right channels over time. Arrows indicate signal bleeding of SR during CB signal detection. (B) Fluorescence profile across the chamber at t = 740 s and t = 960 s. The fluorescence profile of CB shifted as SR fluorescence decreased, indicating the bleeding of the SR fluorescence. }
    \label{bleed}
    \end{figure}

Due to the long tailed excitation spectra of Suoforhodamine 101 (SR) and Fluorescein-dextran (FD), the excitation UV light for Cascade Blue (CB) also excites SR and FD fluorophore.
For example, the SR signal bleeding is evident in Fig. \ref{bleed}; the increase of CB signal on the left channel correlates with the rise of SR signal on the same channel (Fig. \ref{bleed}A, arrows) rather than the CB signal on the right channel (Fig. \ref{bleed}A, blue dashed line). 
Because of this signal bleeding, the steady-state fluorescence profile of CB is altered in the presence of SR (Fig. \ref{bleed}B) or FD.


\section{Carbon source for yeast cell growth in mocrofluidic devices}
We found CO$_2$ production by yeast fermentation disrupts the operation of the platform severely, as the air (CO$_2$) in any parts of the system makes it impossible to control the flow reliably. 
This limits the duration of an experiment. 
To circumvent the problem, we tested various carbon sources to replace glucose in the standard synthetic media for yeast growth, which include glycerol (3\%), potassium acetate (1\%), raffinose (2\%) or maltose (2\%) (see Material and Methods in the main text).
   In normal shaking flask cultures, cell growth is slower with raffinose (doubling time = 3.9 hours) than with glucose (2.4 hours). 
Cell growth with maltose is slower than with those two carbon sources, with doubling time of 5.4 hours.
Other than the slower growth, we found no aberrant morphology of cells with these three carbon sources.
Cell growth in synthetic media with glycerol or potassium acetate is even slower than that in the maltose media (SMM).
 \begin{figure}[t]
    \centering
    \includegraphics[width=0.9\textwidth]{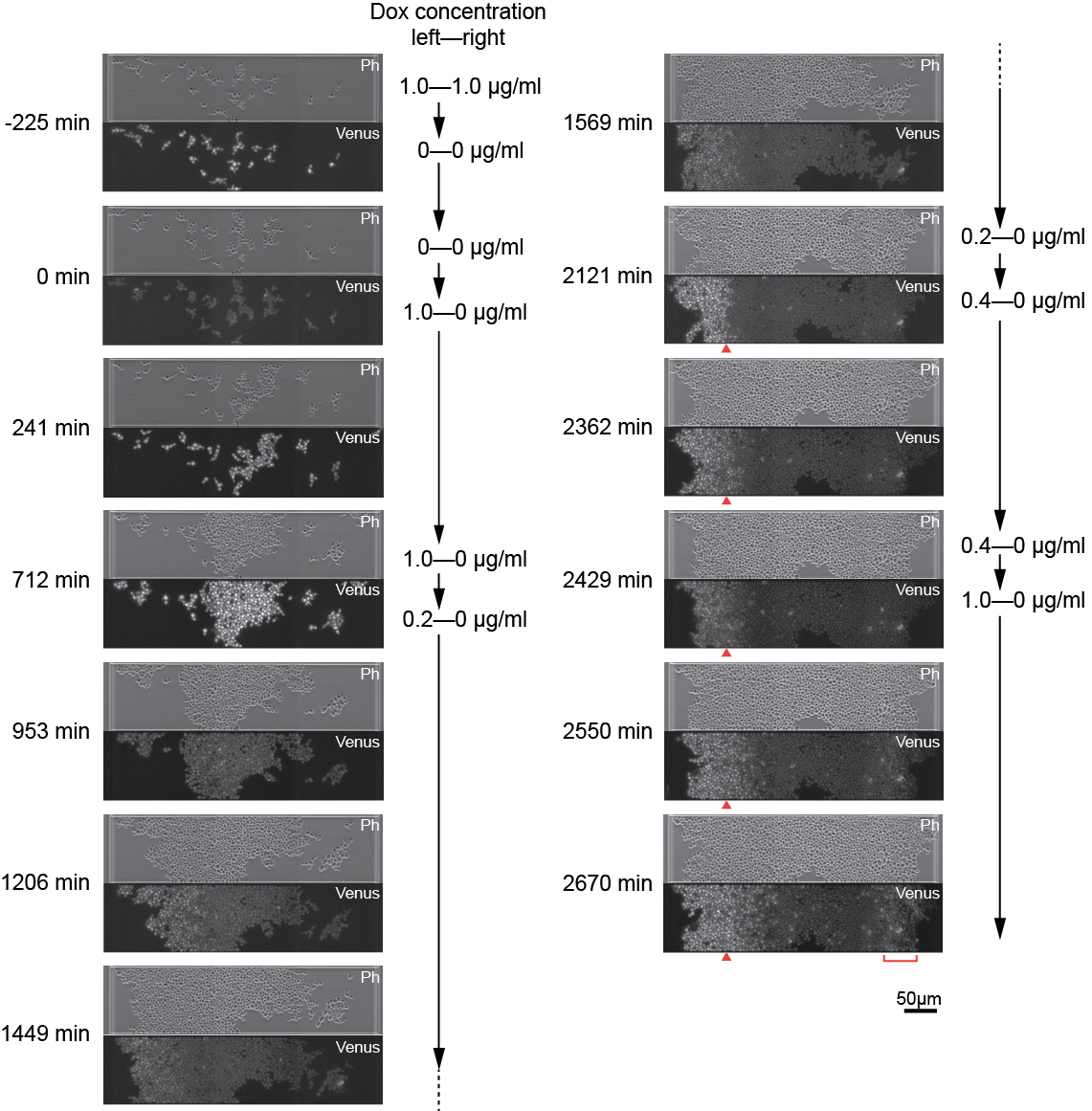}
    \caption{Inducible fluorescent protein reporter expression in yeast cells in a microfluidic chamber. Images (phase contrast (Ph) and Venus) at the indicated time points during the entire course of an experiment are shown. The positions of the expression boundary in the gradient of 0.2-0 $\mu$g/ml are indicated by red arrowheads. Fig. \ref{cell} in the main text is an abridged version of the images shown here. Dox concentrations in the left and right channels are also indicated.}
    \label{venus}
    \end{figure}

    \begin{figure}[ht]
    \centering
    \includegraphics[width=0.5\columnwidth]{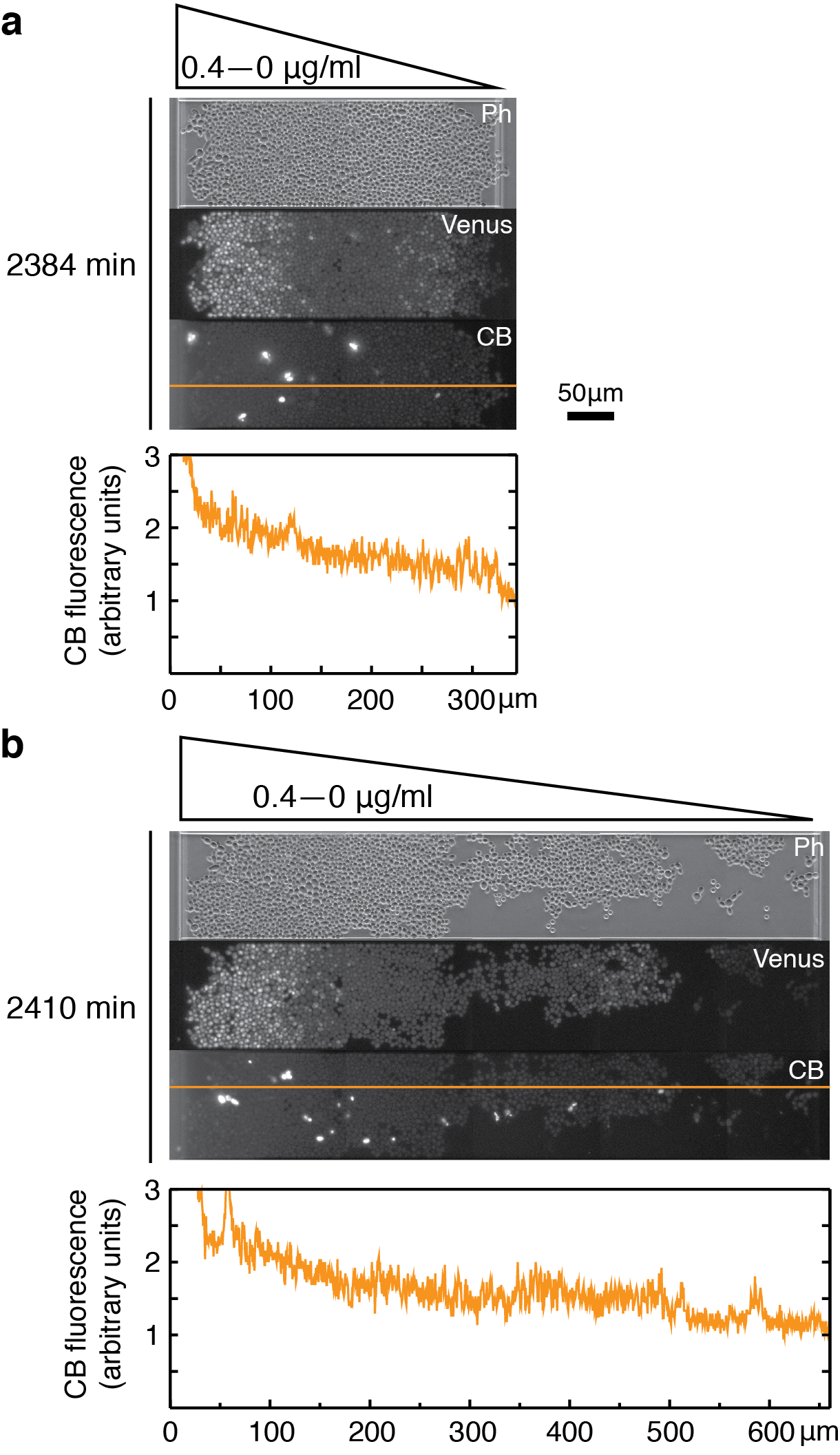}
    \caption{Fluorescent reporter protein expression in yeast in response to the gradient of an inducer. The chamber lengths are 300 $\mu$m (A) and 600 $\mu$m (B). Phase contrast (Ph), Venus, and CB fluorescence images are shown. Bright fluorescence in the image of CB indicates dead cells. Profiles of CB fluorescence (along the yellow line) are also plotted. The image was taken at the indicated time points (the reference time point 0 min is the same as in Fig. \ref{venus}).}
    \label{venus2}
    \end{figure}
We found that yeast cells need to be grown in raffinose media first before being cultured in maltose media; if cells are inoculated into maltose media from a culture of glucose media directly, cells do not grow.
This is perhaps because of a catabolite repression in the presence of glucose. 
Derepression seems to happen in raffinose media, which allows growth with maltose as a sole carbon source.
Although raffinose media extends an operation time for microfluidic experiments, fermentation slowly occurred, leading to an appearance of CO$_2$ bubbles in the system (especially in Tygon tubes).
By contrast, maltose completely eliminated this problem with yeast fermentation in microfluidics.
We therefore decided to use SMM for microfluidic experiments with yeast cells. 
The experiment shown in Fig. \ref{venus} and \ref{venus2} (and Fig. 7 in the main text) were performed using SMM.
 \begin{figure}[hp]
    \centering
    \includegraphics[width=0.6\textwidth]{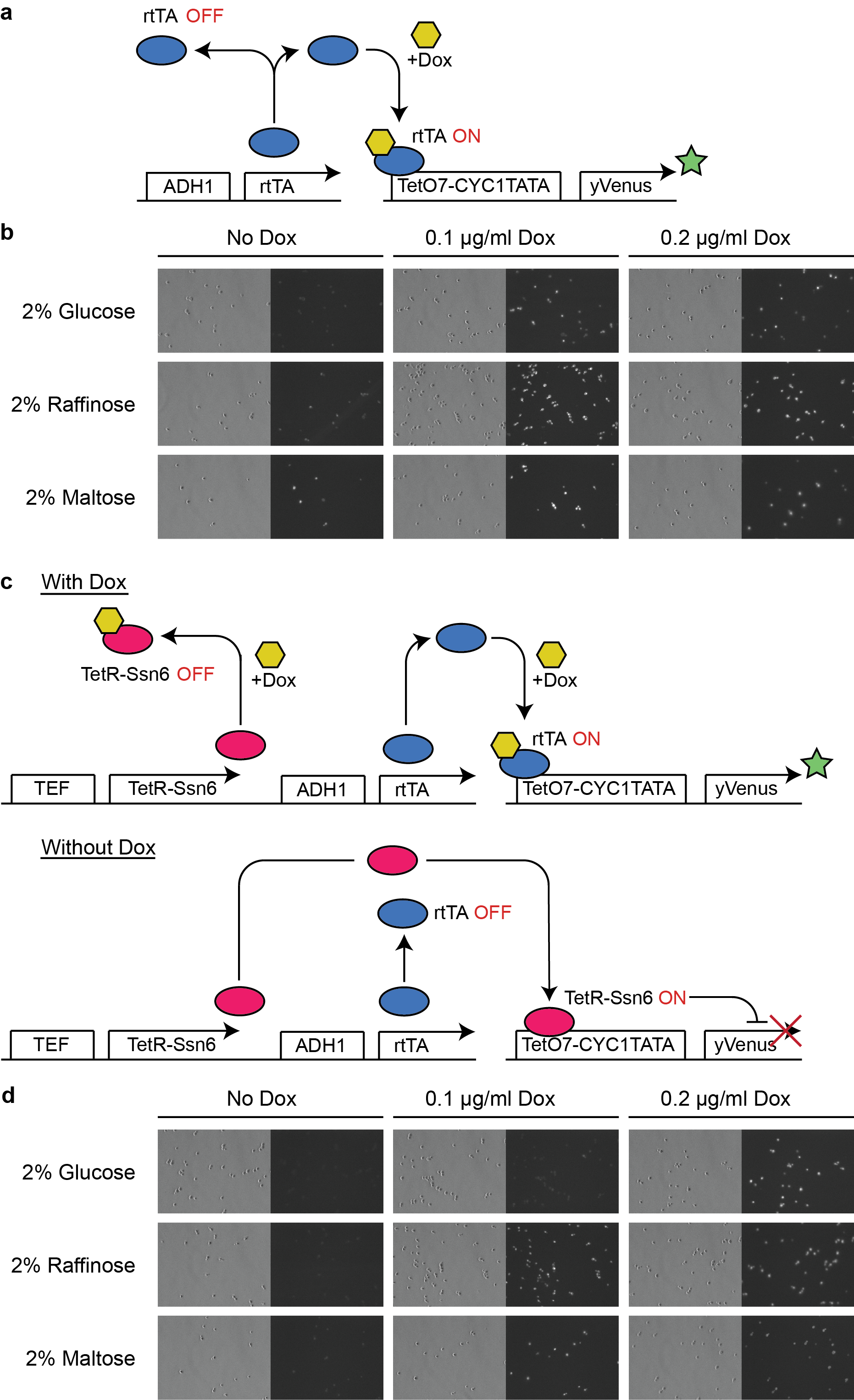}
    \caption{Fluorescent protein reporter expression with various carbon sources. (A, C) Schematics of the inducible expression systems: A, Tet-On system; C, activator/repressor dual system. (B) Expression of yVenus reporter fluorescent protein by the Tet-On system in the absence or presence of doxycycline (Dox). (D) Induction of the reporter gene expression by the dual system in the absence or presence of Dox. Note the background expression of the reporter is suppressed by TetR-Ssn6 in the absence of Dox. In B and D, each pair of panels are phase contrast (left panel) and fluorescence (yVenus, right panel) images. Keys: ADH1 and TEF, constitutively active promoters; rtTA, Dox-inducible transcriptional activator; TetR-Ssn6, a synthetic transcriptional repressor; TetO7-CYC1TATA, the inducible promoter of Tet-On system.}
    \label{reporter}
    \end{figure}
We then examined an inducible fluorescent protein expression by a standard Tet-On system illustrated in Fig. \ref{reporter}A.
Unexpectedly, we observed a high background expression of the reporter yVenus in the absence of the inducer doxycycline (Dox) when cells were cultured in media with raffinose or maltose (but not in glucose) (Fig. \ref{reporter}B).
This makes it difficult to distinguish the induced state from the non-induced state.
We therefore adopted an activator/repressor dual system \cite{Belli:1998qf} illustrated in Fig. \ref{reporter}C.
This system suppressed the background expression of the reporter gene in all three media tested in the absence of Dox (Fig. \ref{reporter}D).
In SMM (and in raffinose media SMR), 0.1 $\mu$g/ml of Dox was sufficient to induce the expression of yVenus.


\section{Detailed methods for microfluidics}\label{sec:setup}
\subsection{Materials}

\begin{table}[t]
\caption{Tygon tubes.}
\begin{center}
\begin{threeparttable}[t]
\begin{tabular}{p{3cm} p{4cm}}
\toprule
Ports & Length\tnote{1} \\
\midrule
M1-M4 &1.3 m \\
S          & 1.2 m + 10 cm \tnote{2} \\
W         & 1.2 m + 5 cm \tnote{2}\\
C1, C2  & 0.8 m + 10 cm \tnote{2}\\
\bottomrule
\end{tabular}
\begin{tablenotes}
\item [1] Use appropriate lengths according to the microscopy system available.
\item [2] Short tubes are connected between the longer one and the chip.
\end{tablenotes}
\end{threeparttable}
\end{center}
\label{table:tab1}
\end{table}

\begin{table}
\caption{Media and solutions for microfluidic operation.}
\begin{threeparttable}[b]
\centering
\begin{tabular}{llr}
\toprule
Ports/Purpose & Media/Solution & Volume\tnote{1}  \\
\midrule
M1, M3        & SMM + 20 $\mu$M CB + 2 $\mu$g/ml Dox & 10 ml\\
M2, M4         & SMM        & 10 ml \\
Cell loading  & SMM      & 5 ml  \\
Priming/wetting chip       & SMM + 0.1\% BSA   & 5 ml \\
C1, C2, S, W     & Water + 0.01\% BSA  &10 ml \\
\bottomrule
\end{tabular}
\begin{tablenotes}
\item [1] Per port.
\end{tablenotes}
\end{threeparttable}
\label{table:tab2}
\end{table}

For media reservoirs connected to M1-M4, W and S ports, 50 ml luer lock syringes were used. 
For C1 and C2 ports, 20 ml syringes were used. 
Syringe reservoirs were covered by Parafilm (pierced through by a needle) to avoid contamination. 
Syringes with media/solutions were connected to Tygon tubes of appropriate lengths through 25Gx5/8" (0.5 mm x 16 mm) needles (Table~\vref{table:tab1}). 
For S, W, C1 and C2 ports, a short tube was inserted between the chip and the longer tube through needle adaptors. 
This arrangement allows draining and replacing syringe reservoirs if necessary. The needle adaptors were made by cutting 25Gx5/8" needles by pliers and grinding off edges on a sandstone sharpener. 
Media/solutions (Table~\vref{table:tab2}) were filter-sterilised (0.2~$\mu$m pore size). 
The solutions were left in a 30 \celsius~incubator to degas for at least a few hours prior to the experiment.
Note that even a tiny air bubble trapped in needles or tubes may disrupt the pressure balance and make flows unsteady.
Tubes were connected to syringes, filled with media/solution by gravity flow and clamped at the ends to stop the flow until being connected to a microfluidic chip.

\subsection{Setting up microfluidics}
\begin{figure}[p]
    \centering
    \includegraphics[width=0.6\columnwidth]{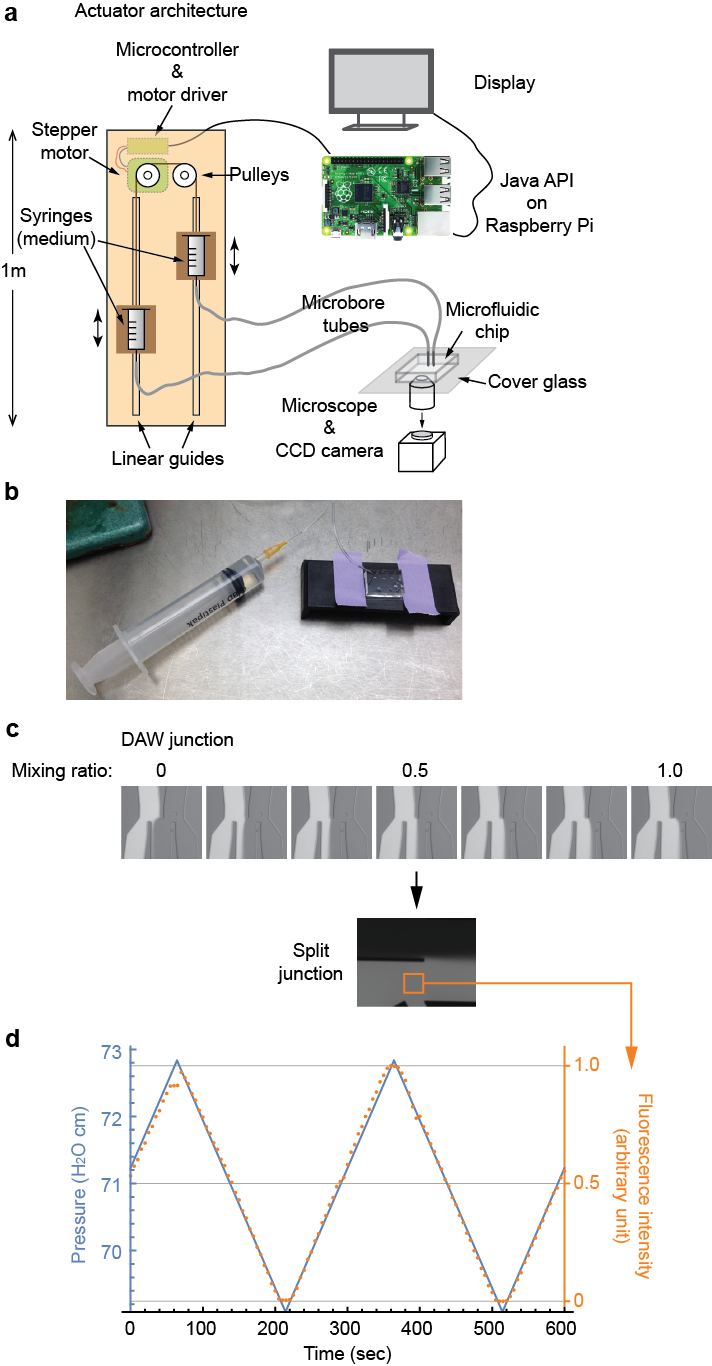}
    \caption{The device setup. (A) Schematic of the system overview. (B) Microfluidic device attached to a syringe for priming/wetting channels. Note the solution flowing out from the ports. (C) Flow of CB at the DAW junction at various mixing ratio. Fluorescence was measured at the rectangular region indicated, and plotted in D (orange dots). (D) The actuator was programmed to produce triangle wave of CB (blue line). The fluorescence intensity fits well to this input over time.}
    \label{device}
    \end{figure}

To prepare the device for operation the procedure below was followed (in a laminar air hood to avoid contamination):
\begin{enumerate}[noitemsep]
\item Affix a microfluidic chip on a solid base (e.g. by Blu-Tack\textsuperscript\textregistered).
\item Draw 5ml SMM+0.1\%~BSA into 20 ml syringe and attach 25Gx5/8" needle. Remove air from the syringe.
\item  Attach a bent needle adaptor to a 10 cm tube and connect the other end to the syringe.
\item  Drain the solution out of the tube/needle then insert the needle adaptor to the W port of the chip.
\item Introduce the solution into the chip by gently pushing the plunger. Leave it to  flow until all the channels are wet and the solution begins flowing out from the other ports (Fig. \ref{device}B). 
\item Attach bent needle adaptors to each tube of the syringe reservoirs prepared above and insert them into the ports in the following order: S $\to$ C1 $\to$ C2 $\to$ M2 $\to$ M4 $\to$ M1 $\to$ M3. BSA in the solution for S, C1 and C2 ports (Table~\ref{table:tab2}) blocks the surface to prevent non-specific binding (adsorption) of CB and cells.
\end{enumerate}

At this stage the microfluidic device was set up on the microscope/imaging system as illustrated in Fig. \ref{device}A, except the syringe reservoir for the port W (still attached to a syringe used to wet the chip). 
Syringe reservoirs attached to M1-M4 were set on the hydrostatic controllers and those attached to C1, C2 and S were set up on a separate clamp stand.
The chip was kept lower than the syringes all the time to avoid air being introduced inside the chip. 
The balances of M1/M3 and M2/M4 ports were also adjusted at this point.
To do this, tubes adjacent to C1 and C2 ports were clamped first and the syringe reservoir heights were fine-tuned by looking at the laminar flow at the split junction (Fig. 1E).

\subsection{Loading yeast cells}

\begin{table}[ht]
\caption{Pressure applied to each port.}
\begin{threeparttable}[t]
\centering
\begin{tabular}{ccc}
\toprule
Operation & Ports & Pressure\tnote{1} \;(height: $cm\, H_2O$)\\
\midrule
Cell loading & M1-M4                            & (closed\tnote{2} ) \\
                    & S                                    & (closed\tnote{2} ) \\
                    & C1, C2                            & 14 \\
                    & W                                    & 14\\
\midrule
Gradient generation & M1-M4               & 71\\
 (cell culture)            & S                        & 37\\
                                 & C1, C2              & (closed\tnote{2} )\\
                                 & W                        & 45\\
\bottomrule
\end{tabular}
\begin{tablenotes}
\item [1] Values are for guidance only; different pressure may be applied according to experimental needs.
\item [2] Ports are closed by clamping Tygon tubes adjacent to the chip.
\end{tablenotes}
\end{threeparttable}
\label{table:tab3}
\end{table}

To load cells into the chip the procedure below was followed:
\begin{enumerate}[noitemsep]
\item Set the pressure to the ports as indicated in Table~\ref{table:tab3}.
\item Draw cell suspension in SMM + 0.01\%~BSA into 20 ml syringe, attach 25Gx5/8" needle and remove air. Do not leave cells for long in the syringe.
\item Remove the syringe connected to the port W and let the solution drain. Attach the syringe of cells to the port through 5 cm tube.
\item Clamp tubes adjacent to M1-M4 and S ports to avoid backward flow during cell injection.
\item Inject cells into the chip. Do not push the plunger too hard. It may take several minutes until cells reach the chip.
\item Once enough cells are injected (which flow towards C1 and C2 ports), remove the syringe and replace it with the syringe tube connected to the syringe reservoir with water + 0.01\% BSA.
\item Trap cells in the diffusion chambers by changing the heights of C1/C2 syringe reservoirs.
\item Once cells are trapped, level the heights of C1/C2 syringe reservoirs (i.e. almost no flow in cell traps) and adjust their position higher than W syringe reservoir to flush cells at the C1 and C2 ports.
\item Once C1 and C2 ports are (almost) clear of cells, clamp W port to stop flow.
\item Clamp both tubes adjacent to the C1 and C2 ports (clamp a few places to stop flow completely), disconnect the longer tubes at the needle adaptors.
\item Remove clamps of M1-M4 and shunt ports. 
\item Remove clamp from waste port to start flow.
\item Set height of syringe reservoirs as suggested in Table~\ref{table:tab3}.
\item Adjust the balance of left and right flows by looking at the laminar flow at the split junction again. Also adjust the balance by looking at loose cells in the chambers by fine-tuning the M1-M4 syringe reservoir heights.

\end{enumerate}

\bibliography{MF}

\end{document}